\documentclass[journal]{IEEEtran}
\usepackage[linesnumbered, ruled]{algorithm2e}
\usepackage{amsmath}
\usepackage{amssymb}
\usepackage{amsfonts}
\usepackage{lipsum}
\usepackage{amsthm}
\usepackage{bm}
\usepackage{color}
\usepackage{cite}
\usepackage{balance}

\usepackage{psfrag}
\usepackage{pgf,tikz} 
\usepackage[hidelinks]{hyperref}
\usepackage{relsize}
\usepackage{graphicx}
\usepackage{epstopdf}
\usepackage{epsfig}
\usepackage{ifpdf}
\usepackage{algorithmic}
\usepackage{url}
\usepackage{array}
\usepackage{subfigure}

\newcommand{\eqdef}{\mathrel{\mathop:}=}
\usepackage{fixltx2e}
\usepackage{stfloats}
\begin{document}
\title{Deep Learning-Enabled Text Semantic Communication under Interference: \\ An Empirical Study}

\author{Tilahun~M.~Getu,~\IEEEmembership{Member,~IEEE,}~Georges~Kaddoum,~\IEEEmembership{Senior Member,~IEEE,}~and~Mehdi~Bennis,~\IEEEmembership{Fellow,~IEEE}%

\thanks{\IEEEcompsocthanksitem T. M. Getu is with the Electrical Engineering Department, \'Ecole de Technologie Sup\'erieure (ETS), Montr\'eal, QC H3C 1K3, Canada (e-mail: tilahun-melkamu.getu.1@ ens.etsmtl.ca).}

\thanks{\IEEEcompsocthanksitem G. Kaddoum is with the Electrical Engineering Department, \'Ecole de Technologie Sup\'erieure (ETS), Montr\'eal, QC H3C 1K3, Canada, and the Cyber Security Systems and Applied AI Research Center, Lebanese American University, Beirut, Lebanon (e-mail: georges.kaddoum@etsmtl.ca).}

\thanks{\IEEEcompsocthanksitem M. Bennis is with the Centre for Wireless Communications, University of Oulu, 90570 Oulu, Finland (e-mail: mehdi.bennis@oulu.fi).}

\thanks{\IEEEcompsocthanksitem The previous research that has inspired this work was supported by the U.S. Department of Commerce and its agency NIST.}

}    % DeepSC, the need for theory, and our computer experiments
\maketitle

% \textcolor{cyan}{

\begin{abstract}
At the confluence of 6G, deep learning (DL), and natural language processing (NLP), DL-enabled text semantic communication (SemCom) has emerged as a 6G enabler since it minimizes bandwidth consumption, transmission delay, and power usage. Among existing text SemCom techniques, a popular text SemCom scheme -- that can reliably transmit semantic information in the low signal-to-noise ratio (SNR) regimes -- is \textit{DeepSC}, whose fundamental asymptotic performance limits under radio frequency interference (RFI) were accurately predicted by our recently developed theory \cite{Getu_TWC'24}. Although our theory was corroborated by simulations, trained deep networks can defy classical statistical wisdom, calling for extensive computer experiments. This empirical work thus follows using the training, validation, and testing sets \textit{tokenized and vectorized} from the \textit{Proceedings of the European Parliament (Europarl)} dataset. Specifically, we train the DeepSC architecture in Keras 2.9 with TensorFlow 2.9 as a backend and test it under Gaussian multi-interferer RFI received over Rayleigh fading channels. Our testing results corroborate that DeepSC produces semantically irrelevant sentences under huge Gaussian RFI emitters, validating our theory. Therefore, a fundamental 6G design paradigm for \textit{interference-resistant and robust SemCom} (IR$^2$ SemCom) is needed. 
                                      
\end{abstract}

\IEEEpeerreviewmaketitle
\section{Introduction}
\label{sec: SS_Intro} 
Over the last decade, deep learning (DL) \cite{YYBGH_15} has been propelling the rise of artificial intelligence (AI) \cite{Russel_AI_Book_18}, in general, and machine learning (ML), in particular, leading to numerous breakthroughs in various fields of science, technology, engineering, and mathematics. In computer science, for instance, DL has led to numerous remarkable results in -- among other areas -- image recognition \cite{He_ResNet_CVPR_16}, object detection \cite{LLL_DL_for_GOB_19}, speech recognition \cite{Hinton_Speech_Recognition'12}, and natural language processing (NLP) \cite{Torfi_NLP_Survey'20,Trends_in_DL-Based_NLP'18}. Due to the rise of DL in NLP, statistical machine translation (SMT) has been remarkably superseded by neural machine translation (NMT) \cite{NMT_Review'20}.     

DL has spawned the depth and breadth of sixth-generation (6G) research \cite{Vis_6G_VT_19,Alwis_6G_Frontiers'21,Alsabah_6G_Wireless_Commun_Network'21,Letaief_6G_Roadmap_2019} toward ultra-reliable ubiquitous communication, networking, and sensing \cite{URLLC_Bennis_18}. With a potential to materialize such 6G services, DL-enabled semantic communication (SemCom) \cite{Xie_DeepSC_TSP'21,Getu2023_tutorialcumsurvey,Sem_Empowered_Commun'22} has emerged to realize Weaver's 1949 vision \cite[Ch. 1]{Shannon_Weaver_Math_Theory_Commun'49} of a \textit{meaning-centric} communication, while minimizing transmission delay, bandwidth consumption, and power usage. Specifically, DL has spurred the development of numerous DL-enabled SemCom techniques in text \cite{Xie_DeepSC_TSP'21}, speech \cite{Weng_SemCom_Sys_Speech_Trans'21}, image \cite{Huang_Toward_SemCom'23}, and video \cite{Huang_IS-SemCom'22} domains. Among the existing text SemCom techniques \cite{Getu2023_tutorialcumsurvey,Getu2023_techrxiv_23}, \textit{DeepSC} is a popular SemCom technique that can reliably transmit semantic information in the low signal-to-noise ratio (SNR) regimes. However, DeepSC can be severely impacted by \textit{semantic noise} caused by radio frequency interference (RFI) from one or more RFI emitters \cite{Getu_TWC'24}. Hence, the performance quantification of DeepSC under interference informs the design of \textit{interference-resistant and robust (IR$^2$) SemCom systems} \cite{Getu_TWC'24}. 

Toward IR$^2$ SemCom systems, the fundamental asymptotic performance limits of DeepSC were derived by our recent work \cite{Getu_TWC'24}, which theorized that DeepSC produces \textit{semantically irrelevant} sentences when the RFI emitters -- rendering multi-interferer RFI (MI RFI) \cite{Getu_dissertation_19} -- get strong and become enormous. These performance limits were corroborated by Monte Carlo simulations, though trained deep networks can defy classical statistical wisdom \cite{Belkin_PNAS_reconciling_MML'19,Zhang_Understanding'17,Bartlett_DL_Stat_Viewpoint'21,Adlam_Triple_Descent'20}. Accordingly, computer experiments are needed to verify the performance limits predicted by our theory in \cite{Getu_TWC'24}. Meanwhile, there is a lack of previous research\footnote{To the best of our knowledge, there is a lack of studies on the impact of interference on a text SemCom system. However, the authors of \cite{hu2022robust} studied the effect of semantic noise (due to a malicious attacker) on an image SemCom system, and the authors of \cite{Sagduyu2022_SemCom} investigated the impact of wireless attacks also on an image SemCom system.} on the impact of interference on text SemCom systems, as reported by our surveys in \cite{Getu2023_tutorialcumsurvey,Getu2023_techrxiv_23,Getu_Go-SemCom_Survey'24}, justifying the motivation and need for this paper.

Employing a standard SMT and NMT dataset named the \textit{Proceedings of the European Parliament (Europarl)} \cite{Europarl_dataset}, in this paper, we document our extensive computer experiments on the training of DeepSC and testing of its trained models with and without MI RFI. Major contributions of this study are summarized as follows:       
\begin{itemize}
	
	\item  We present a detailed description of DeepSC's training using Keras 2.9 with TensorFlow 2.9 as a backend. 
	
	\item  We test trained DeepSC models with and without MI RFI received over Rayleigh fading channels. 
	
	\item We empirically demonstrate that DeepSC produces semantically irrelevant sentences as the number of Gaussian RFI interferers becomes enormous, confirming our recently developed theory on DeepSC's performance limits.
		
	\item Through detailed step-by-step procedures, we establish a data preprocessing and neural processing standard for DeepSC and DeepSC-inspired SemCom techniques.		
\end{itemize}
Informed by our multidisciplinary theoretical and empirical research on DL, NLP, NMT, and SemCom, in this paper, we document details on the entire step-by-step procedures on the training of DeepSC followed by its testing with and without interference. In doing so, we aim to bridge the existing knowledge gap -- from an implementation viewpoint -- that can hamper the development of many text SemCom techniques. 

The remainder of this paper is organized as follows. Section (Sec.) \ref{sec: training_setup_and_assumptions} presents our training setup and assumptions. Sec. \ref{sec: Data_standardization_tokenization_and_vectorization} details data standardization, tokenization, and vectorization. Sec. \ref{sec: DeepSC_training} documents end-to-end training of DeepSC. Sec. \ref{sec: DeepSC_testing} reports on the testing results with and without MI RFI. Finally, Sec. \ref{sec: summary_and_outlook} provides a concluding summary and research outlook. 
 
\begin{figure*}[!t]
	\centering
	\includegraphics[scale=0.50]{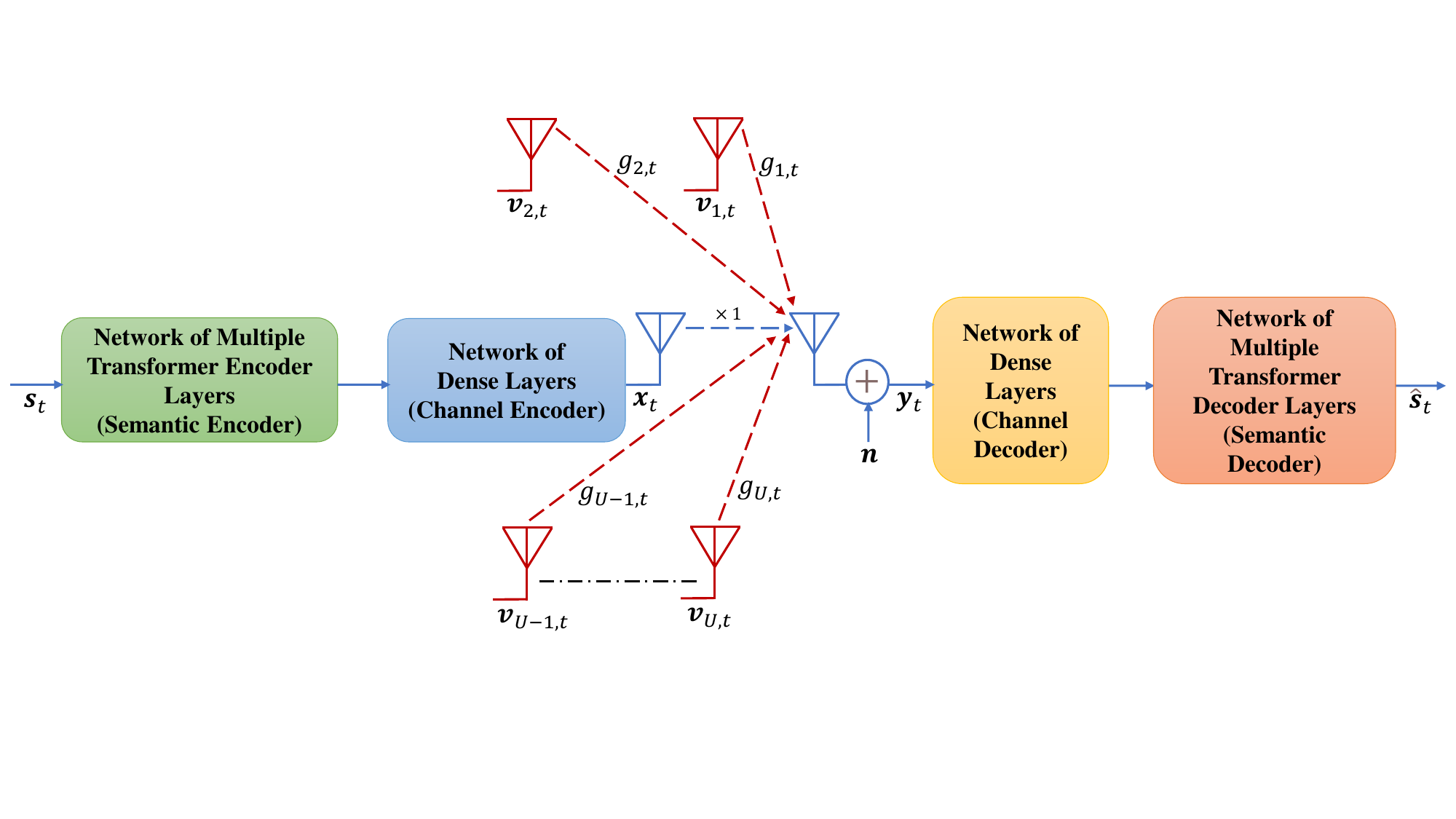}   \\ [-2cm]
	\caption{System setup: During DeepSC training, the $t$-th DeepSC symbol $\bm{x}_t$ is transmitted over an additive white Gaussian noise (AWGN) channel; during DeepSC testing, the DeepSC symbols -- transmitted over an AWGN channel -- are received along with the symbols of time-varying MI RFI from $U$ ($U\geq 2$) Gaussian RFI emitters, whose interference signals are received over Rayleigh fading channels.} 
	\label{fig: DeepSC_System_Model_20220921}
\end{figure*}  
\textit{Notation}: Scalars, vectors, and matrices (also tensors) are represented by italic letters, bold lowercase letters, and bold uppercase letters, respectively. $\mathbb{N}$, $\mathbb{R}$, $\mathbb{R}^n$, $\mathbb{R}^{m \times n}$, and $\mathbb{R}^{m \times n \times p}$ denote the sets of natural numbers, real numbers, $n$-dimensional vectors of real numbers, $m \times n$ real matrices, and $m \times n \times p$ real three-way tensors, respectively. $\eqdef$, $\sim$, $(\cdot)^T$, $\mathbb{P}(\cdot)$, $\| \cdot \|$, $\mathbb{I}\{\cdot\}$, $\bm{0}$, and $\bm{I}_n$ stand for equal by definition, distributed as, transpose, probability, Euclidean norm, an indicator function that returns one when the argument is true and 0 otherwise, a zero vector, and an $n \times n$ identity matrix, respectively. $[n] \eqdef \{1, 2, \ldots, n\}$. For a row vector $\bm{a} \in \mathbb{R}^{1 \times n}$ and a column vector $\bm{b} \in \mathbb{R}^{n}$, their $i$-th elements are denoted by $(\bm{a})_i$ and $(\bm{b})_i$, respectively. The dot product between two conformable vectors $\bm{a}$ and $\bm{b}$ is denoted as $\bm{a}\cdot\bm{b}$. $\mathcal{N}(0, \sigma^2)$ denotes a zero-mean Gaussian distribution with a variance of $\sigma^2$. A vector $\bm{x} \eqdef [ (\bm{x})_1, (\bm{x})_2, \ldots, (\bm{x})_n ] \in \mathbb{R}^n$ is characterized as $\bm{x} \sim \mathcal{N}(\bm{0}, \sigma^2\bm{I}_n)$ if and only if (iff) all its elements are jointly independent and Gaussian, i.e., $(\bm{x})_i \sim  \mathcal{N}(0, \sigma^2)$ $\forall i \in [n]$.  

\section{Training Setup and Assumptions} 
\label{sec: training_setup_and_assumptions}
Our theoretical work in \cite{Getu_TWC'24} quantified the performance of a text SemCom system named DeepSC when it suffers from RFI emitted by one or more single-antenna RFI emitters. To validate our performance quantification, we conduct extensive end-to-end training of DeepSC -- as shown in Fig. \ref{fig: DeepSC_experiment_model_20230715} -- followed by its testing with and without MI RFI. For this setup, we outline below the system setup and assumptions of our computer experiments, as shown in Fig. \ref{fig: DeepSC_System_Model_20220921}.

Let $\bm{s}_t \eqdef [w_{1,t}, w_{2,t}, \ldots, w_{L,t} ]$ be a sentence of $L$ words to be transmitted using DeepSC during the $t$-th time slot.\footnote{Without loss of generality, we assume a time-slotted system whose one time slot equals $KL$ times the duration of one semantic symbol, as per (\ref{x_vec_defn}).} The DeepSC transmitter first feeds $\bm{s}_t$ to a semantic encoder whose outputs are then fed to a channel encoder to produce the $t$-th \textit{DeepSC symbol} $\bm{x}_t$ that is given by \cite{Xie_DeepSC_TSP'21}
\begin{equation}
\label{x_vec_defn}
\bm{x}_t \eqdef C_{\bm{\alpha}} (S_{\bm{\beta}}(\bm{s}_t))   \in \mathbb{R}^{1 \times KL}, 
\end{equation}
where $S_{\bm{\beta}}(\cdot)$ and $C_{\bm{\alpha}} (\cdot)$ denote the semantic encoder and channel encoder networks with parameter sets $\bm{\beta}$ and $\bm{\alpha}$, respectively; $K$ stands for the average number of semantic symbols per a word in $\bm{s}_t$; and $\bm{x}_t^T \in \mathbb{R}^{KL}$, since we consider real inputs without loss of generality. The $t$-th DeepSC symbol is transmitted through an AWGN channel -- which we assume without loss of generality -- that yields the $t$-th received DeepSC signal $\bm{y}_t$ equated as  
\begin{equation}
\label{y_vec_defn}
\bm{y}_t \eqdef \bm{x}_t + \bm{n}   \in \mathbb{R}^{1 \times KL}, 
\end{equation}
where $\bm{n}$ represents the contaminating AWGN characterized as $\bm{n} \sim \mathcal{N}(\bm{0}, \sigma^2\bm{I}_{KL})$. The $t$-th received DeepSC signal $\bm{y}_t$ goes through the channel decoder $C_{\bm{\delta}} (\cdot)$ whose outputs are fed to the semantic decoder $S_{\bm{\theta}} (\cdot)$ to produce the $t$-th recovered training sentence $\hat{\bm{s}}_t$ given by \cite{Xie_DeepSC_TSP'21}
\begin{equation}
\label{s_hat_vec_defn}
\hat{\bm{s}}_t \eqdef S_{\bm{\theta}} (C_{\bm{\delta}}(\bm{y}_t)), 
\end{equation}
where the channel decoder $C_{\bm{\delta}} (\cdot)$ and the semantic decoder $S_{\bm{\theta}} (\cdot)$ have parameter sets $\bm{\delta}$ and $\bm{\theta}$, respectively.

End-to-end training of DeepSC is equivalent to determining the (nearly) optimum semantic encoder, channel encoder, channel decoder, and semantic decoder that minimize the \textit{semantic discrepancy} between the $t$-th recovered and transmitted sentences $\hat{\bm{s}}_t$ and $\bm{s}_t$, $\forall t\in \mathbb{N}$, respectively. This training takes $\bm{s}_t$ as the $t$-th input to the DeepSC architecture whose recovered sentence $\hat{\bm{s}}_t$ is compared with respect to (w.r.t.) label $\bm{s}_t$ so that end-to-end training of DeepSC resumes using the back-propagation algorithm (BackProp). However, this supervised learning problem cannot be formulated as a regression AI/ML problem, since neither computers nor deep networks can understand strings/sentences, but numbers.\footnote{If computers and deep networks were able to understand strings/sentences, they would be the training inputs and labels of a regression AI/ML problem on NMT and text SemCom.} To overcome this limitation, one would \textit{vectorize} each sentence into a sequence of integers by converting each word into integers (i.e., \textit{tokenize}), which are then fed to the DeepSC network that would be trained using \textit{categorical cross-entropy} to solve a multi-level classification AI/ML problem by learning the probability of each word in a sentence, while taking into account the dictionary size (vocabulary size) of a given dataset. 

The probability of each word can be learned via a DeepSC training w.r.t. training labels that are the \textit{one-hot encoded} \cite{Keras_CategoryEncoding_Layer} versions of the tokenized words of $\bm{s}_t$ -- $\forall t \in \mathbb{N}$, leading to our data standardization, tokenization, and vectorization.   
 
\section{Data Standardization, Tokenization, and Vectorization}	
\label{sec: Data_standardization_tokenization_and_vectorization}   
The Europarl parallel corpora are extracted from the European Parliament's proceedings and include versions in 21 European languages \cite{Europarl_dataset}. Europarl \cite{Europarl_dataset} is a standard dataset that has been deployed for SMT and NMT \cite{Brownlee_How_to_Prepare}, as well as the training and testing of DL-based text SemCom systems.\footnote{Europarl was used to train many text SemCom systems \cite[Sec. III]{Getu2023_tutorialcumsurvey}.} Accordingly, we also adopt Europarl in our training and testing of DeepSC, and standardize it as described below. 

From Europarl's 20 parallel corpora in \cite{Europarl_dataset}, we first download the German-English parallel corpus (193 MB) comprising 1,920,209 ($\approx$ 2 million) sentences and 47,818,827 ($\approx$ 48 million) English words. We then \textit{unzip} the downloaded corpus and upload the English corpus (named \textquotedblleft europarl-v7.de-en.en'') into our working directories at the graphics processing unit (GPU) clusters of the Digital Research Alliance of Canada \cite{DRAC_address} named \textit{Béluga} \cite{DRAC_Beluga} and \textit{Graham} \cite{DRAC_Graham}. Our uploaded English corpus needs to be standardized (or \textit{cleaned}) since all non-printable characters, punctuation characters, and words with non-alphabetic characters have to be removed \cite{Brownlee_How_to_Develop,Brownlee_How_to_Prepare}, as implemented below. 

\subsection{Data Standardization}
\label{subsec: data_standardization}
We begin our data standardization by uploading the Europarl text into memory. On our working directory in both Béluga and Graham, we first load the Europarl English document into memory by opening it as a read-only file, reading all its text, closing the file, and returning it as a \textit{blob of text}. We then split the loaded text into sentences by splitting it on new line characters. We then clean each of these sentences by normalizing it to Unicode characters, tokenizing it on white space, converting it to lowercase, removing punctuation and non-printable characters from each of its tokens, removing its tokens with numbers in them, and, finally, storing it as a string. We then return a list of clean sentences that we save to a file, which we henceforward refer to as \textit{our saved Europarl document of clean sentences}. This document can comprise many less frequent words -- which hardly help DeepSC to learn a text SemCom system efficiently -- that only increase the vocabulary size to the extent that an \textit{out-of-memory error (OOM error)} is triggered by Béluga and Graham, justifying the need for vocabulary size reduction, as implemented below.

\subsection{Data Vocabulary Reduction}
\label{subsec: data_vocabulary_reduction}
Our implementation reduces the vocabulary of our saved Europarl document of clean sentences by marking all the \textit{out-of-vocabulary} words with a special NMT token named \textquotedblleft unk''. Accordingly, we first load our saved Europarl document of clean sentences; from these loaded clean sentences, we create a vocabulary; from this vocabulary, we remove all words that have an occurrence below a minimum specified threshold and generate our trimmed vocabulary; by taking our trimmed vocabulary and the loaded clean sentences as inputs, we create \textit{our Europarl document of clean sentences with reduced vocabulary} after removing all words that are not in our trimmed vocabulary and marking their removal with \textquotedblleft unk\textquotedblright -- similar to the work in \cite{Brownlee_How_to_Prepare}. We save this document, which we use to prepare our training, validation, and testing sets.

Having trimmed words that appear less than 20 times in our saved Europarl document of clean sentences, we reduced the vocabulary size from 102,917 to 22,897: Our saved Europarl document of clean sentences with reduced vocabulary has thus a dictionary size of 22,897 words. By loading this document with 1,920,209 sentences, we saved the first 1,900,000 sentences as the overall dataset sentences. From these sentences, we saved the first 100,000, the next 1,500,000, and the last 300,000 sentences as testing sentences, training sentences, and validation sentences, respectively. Using these three sets of sentences, we prepare our testing, training, and validation sets, respectively, as detailed below.    

\subsection{Preparation of the Testing, Training, and Validation Sets} 
\label{subsec: preparation_of_datasets}
In preparing our testing, training, and validation sets, \textit{data tokenization} and \textit{data vectorization} were needed.
\subsubsection{Data Tokenization}
\label{subsec: data_tokenization}
As neither DL networks nor computers take strings as input, each word of the overall dataset sentences must be encoded into numbers, and all its sentences must be converted to sequences of numbers. This is known as data tokenization (or \textit{text tokenization}) in NMT. In our text tokenization, we exploited the Keras \textit{Tokenizer} class \cite{TensorFlow_Tokenizer} to map words to integers, and fed the 1.9 million sentences to a Keras function named \texttt{fit$\_$on$\_$texts()} \cite{TensorFlow_Tokenizer}. The returned tokenizer was then fed to the data vectorizer for data vectorization (or \textit{data encoding}). 

\subsubsection{Data Vectorization}
\label{subsec: data_vectorization}
Regarding the testing, training, and validation sets, every input and output sequence should be encoded to integers and padded/truncated to the optimal\footnote{A number of NMT works, such as the work in \cite{Brownlee_How_to_Develop}, encoded (to integers) and padded each input sequence to the length of the longest sequence in a list of sentences. This strategy triggered an OOM error in both Béluga and Graham. In our DeepSC training, we found $L=30$ to be an optimal length that does not cause an OOM error. Hence, we truncate and zero pad sequences with a length greater than and less than 30, respectively.} length $L$, which is also the number of input neurons -- of a DL network such as DeepSC -- that would not trigger an OOM error. This is data vectorization that we have accomplished using an encoding function taking the tokenizer of Sec. \ref{subsec: data_tokenization}, length $L=30$, and the testing, training, or validation sentences. Specifically, this encoding function executes the following actions on the testing, training, and validation sentences:
\begin{enumerate}
	\item It first transforms each sentence of the testing, training, and validation sentences to a sequence of integers by employing the tokenizer of Sec. \ref{subsec: data_tokenization} and the Keras function \texttt{texts$\_$to$\_$sequences()} \cite{TensorFlow_Tokenizer}. 
	\item It then \textit{vectorizes} each sequence of integers to an integer sequence of (the same) length $L=30$ by post padding with zeros or truncating using the Keras function \texttt{tf.keras.utils.pad$\_$sequences()} \cite{TensorFlow_pad_sequences}.
\end{enumerate}
Each token of the produced sequences (of length $L=30$) needs to be labeled, as highlighted below.

\subsubsection{Data Labeling}
\label{subsec: data_labeling}
\begin{figure*}[!t]
	\centering
	\includegraphics[scale=0.46]{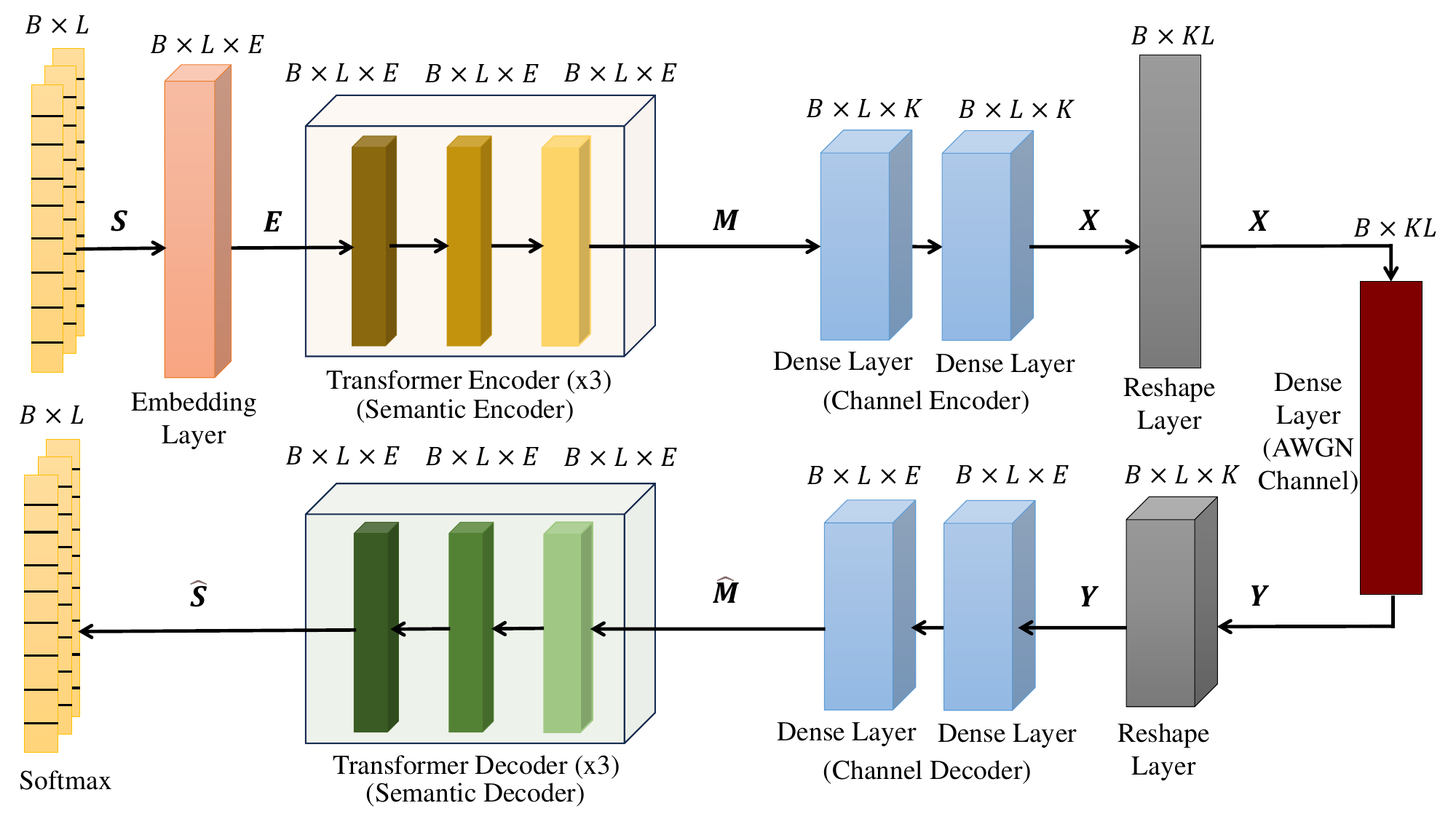}   \\
	\caption{The DeepSC architecture (with real inputs) under our training and testing. (Hyper)parameters: $B$ -- batch size; $L$ -- the number of words per a transmitted sentence (i.e., a transmitted sentence length); $E$ -- embedding dimension (the output dimension of an embedding layer); $K$ -- the average number of semantic symbols per word in a given transmitted sentence.} 
	\label{fig: DeepSC_experiment_model_20230715}
\end{figure*}

Since the DeepSC model would be trained to learn the probability of each word (via its assigned unique token) w.r.t. its \texttt{softmax} prediction layer, each token of a data vectorized sequence has to be \textit{one-hot encoded}. To accomplish this, we exploit the Keras function \texttt{tf.keras.utils.to$\_$categorical()} \cite{TensorFlow_to_categorical} that we apply to each data vectorized sequence of Sec. \ref{subsec: data_vectorization}. This one-hot encoding is also fed the number of classes equal to the vocabulary size of our tokenizer per Sec. \ref{subsec: data_tokenization}. The vocabulary size of our tokenizer is equal to the number of unique word indices in our Sec. \ref{subsec: data_tokenization}'s tokenizer plus one,\footnote{In our DeepSC training and testing, the vocabulary size was equal to 22,899, which was calculated after vocabulary reduction per Sec. \ref{subsec: data_vocabulary_reduction}.} which is also the number of output neurons in the \texttt{softmax} prediction layer of the DeepSC architecture shown in Fig. \ref{fig: DeepSC_experiment_model_20230715}.   

According to the data tokenization, data vectorization, and data labeling of Secs. \ref{subsec: data_tokenization}, \ref{subsec: data_vectorization}, and \ref{subsec: data_labeling}, we prepare our training, validation, and testing sets. When it comes to our testing set, we do not need testing labels since we assess the performance of our trained DeepSC model -- without and with RFI -- using a SemCom performance assessment metric named \textit{sentence similarity} \cite{Getu_IEEE_Access'23}. Hence, we only prepare the testing inputs by loading the saved 100,000 testing sentences (from Sec. \ref{subsec: data_vocabulary_reduction}) that we tokenized and vectorized per Secs. \ref{subsec: data_tokenization} and \ref{subsec: data_vectorization}, respectively. This produces testing inputs of $100,000 \times 30$ (non-negative) integers. Meanwhile, we prepare our training set by implementing the following four steps in Keras with TensorFlow as a backend:
\begin{enumerate}
	\item We first load our saved 1.5 million training sentences (from Sec. \ref{subsec: data_vocabulary_reduction}) to our working directories in Béluga and Graham.
	
	\item We then tokenize and vectorize -- per Secs. \ref{subsec: data_tokenization} and \ref{subsec: data_vectorization}, respectively -- each training sentence  to produce our training inputs of $1,500,000 \times 30$ (non-negative) integers.  
	
	\item We then feed the produced training inputs to a Keras generator function -- implemented using \texttt{tf.keras.utils.Sequence} \cite{TensorFlow_utils_sequence} along with the data labeling per Sec. \ref{subsec: data_labeling} -- that also takes a batch size $B=50$ and a vocabulary size of 22,899.
	
	\item Our coded Keras generator function\footnote{Our coded Keras generator function randomly shuffle our training and validation sets -- between epochs -- on every epoch end. We implement the data labeling of Sec. \ref{subsec: data_labeling} inside our coded Keras generator function to overcome an OOM error, which we came across in both Béluga and Graham.} then returns our training inputs of $50 \times 30$ (non-negative) integers and training labels of $50 \times 30 \times 22,899$ binary integers (zeros or ones) for every training batch.
\end{enumerate}

Similarly, we prepare our validation set by executing the following four steps in Keras with TensorFlow as a backend:
\begin{enumerate}
	\item We first load our saved 300,000 validation sentences (from Sec. \ref{subsec: data_vocabulary_reduction}) to our working directories in Béluga and Graham.
	
	\item We then tokenize and vectorize -- per Secs. \ref{subsec: data_tokenization} and \ref{subsec: data_vectorization}, respectively -- each validation sentence to yield our validation inputs of 300,000 $\times$ 30 (non-negative) integers. 
	
	\item We then feed the validation inputs to the aforementioned Keras generator function which also takes a batch size of $50$ and a vocabulary size of 22,899, among its inputs.
	
	\item Our coded Keras generator function then returns our validation inputs of $50 \times 30$ (non-negative) integers and validation labels of $50 \times 30 \times 22,899$ binary integers for every validation batch.
\end{enumerate}
On being returned by our Keras generator, our prepared training and validation sets are then fed to the Keras \texttt{model.fit$\_$generator()} \cite{Keras_Model_Training_APIs} training function after defining and compiling the DeepSC model, as detailed below.

\section{End-to-End Training of DeepSC}
\label{sec: DeepSC_training} 
In this section, we present the end-to-end training architecture of DeepSC, the DeepSC model definition, and the training results of DeepSC that we obtained using \texttt{Adam} \cite{Keras_Adam_Opt}, beginning with DeepSC's end-to-end training architecture.  
\subsection{The End-to-End Training Architecture of DeepSC}
\label{subsec: DeepSC_architecture}
\begin{figure*}[!htb]
	\centering
	\includegraphics[scale=0.46]{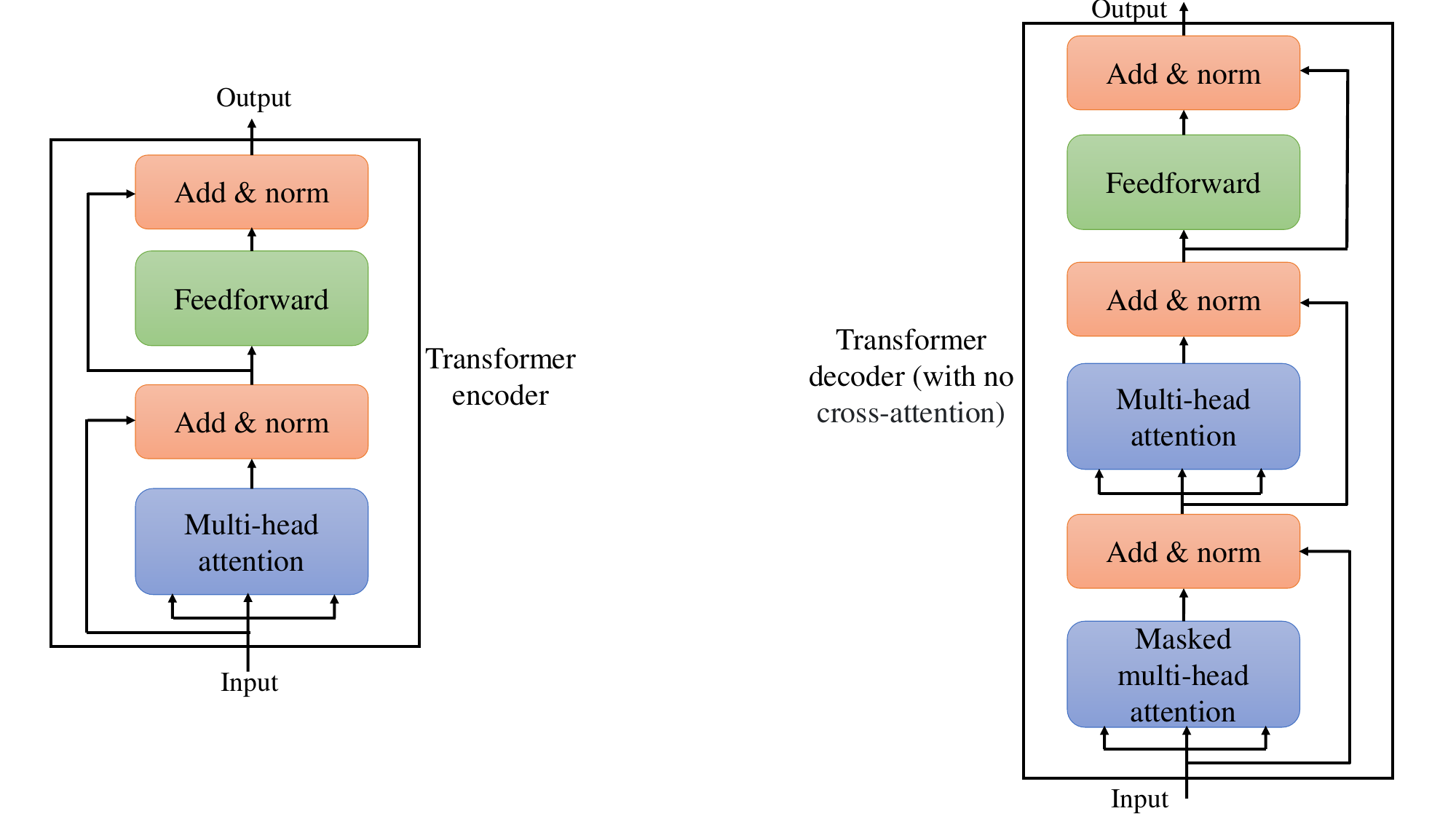}   \\
	\caption{Transformer encoder and Transformer decoder (with no cross-attention) \cite{Vaswani2023_attention}.} 
	\label{fig: Trans_encoder_decoder_20230803}
\end{figure*}

The DeepSC architecture (with real inputs) that we employ in our end-to-end training of DeepSC is shown in Fig. \ref{fig: DeepSC_experiment_model_20230715},\footnote{The DeepSC authors \cite{Xie_DeepSC_TSP'21} first trained the \textit{MINE (mutual information neural estimation) network} from \cite{Belghazi'21_MINE} to maximize the mutual information $I(\bm{x}_t; \bm{y}_t)$. They then used the loss of this network to tweak the loss of the DeepSC architecture -- shown in Fig. \ref{fig: DeepSC_experiment_model_20230715} -- during its training, which they believe maximizes the achieved data rate \cite{Xie_DeepSC_TSP'21}. However, since our problem is assessing the impact of RFI on \textit{sentence similarity}, we discard training the MINE network and directly train the DeepSC architecture.} which shows the forward-propagation during \textit{one batch} of training inputs with a batch size $B$; a matrix $\bm{S} \in  \mathbb{R}^{B \times L}$ of training inputs is fed to an embedding layer that turns (positive-integer) indexes into dense vectors of fixed size \cite{Keras_Embedding_Layer}, yielding an embedding tensor $\bm{E} \in \mathbb{R}^{B \times L \times E} $. The embedding tensor $\bm{E}$ is fed to the semantic encoder made of three cascaded Transformer encoders (see Fig. \ref{fig: Trans_encoder_decoder_20230803}) that produce rich representations (for every embedded word vector) which constitute an output tensor $\bm{M} \in \mathbb{R}^{B \times L \times E}$. Being a tensor of semantically-encoded symbols, the output tensor $\bm{M}$ is fed to a channel encoder\footnote{Note that the channel encoder/decoder can be composed of numerous cascaded \texttt{Dense} layers, which are deep networks in their own right. However, as demonstrated by our computer experiments, deeper and (possibly) overfitted architectures led to \textit{gradient explosion} and/or \textit{gradient vanishing}.} made of two cascaded \texttt{Dense} layers \cite{Keras_Dense_Layer} that give an output tensor $\bm{X} \in \mathbb{R}^{B \times L \times K	}$. The channel-encoded semantic symbols are then reshaped to produce $\bm{X} \in \mathbb{R}^{B \times KL}$, which are transmitted through an AWGN channel modeled by a linear \texttt{Dense} layer.

The AWGN channel contaminates its input $\bm{X} \in \mathbb{R}^{B \times KL}$, and produces an output reshaped into a three-way tensor $\bm{Y} \in \mathbb{R}^{B \times L \times K}$. Tensor $\bm{Y}$ is inputted to a channel decoder composed of two cascaded \texttt{Dense} layers that yield a tensor $\hat{\bm{M}} \in \mathbb{R}^{B \times L \times E}$, which is a tensor of recovered semantically-encoded symbols. Tensor $\hat{\bm{M}} \in \mathbb{R}^{B \times L \times E}$ is then fed to a semantic decoder composed of three cascaded Transformer decoders -- with no cross-attention as in Fig. \ref{fig: Trans_encoder_decoder_20230803} -- whose outputs are fed to a \texttt{softmax} prediction layer to produce a matrix $\hat{\bm{S}} \in  \mathbb{R}^{B \times L}$, which is a matrix of recovered sentences. In line with Figs. \ref{fig: DeepSC_experiment_model_20230715} and \ref{fig: Trans_encoder_decoder_20230803}, we now present the DeepSC model definition. 

\subsection{Model Definition of DeepSC}
\label{subsec: DeepSC_model_definition}
Without loss of generality and to prevent an OOM error in both Béluga and Graham, we defined two DeepSC models for our training. These models are a narrow model named \textit{narrow DeepSC} and a relatively wide model named \textit{relatively wide DeepSC} that are parameterized by $(K, H, V, E, L)$, where $(K, E, L)$ are parameters defined in the caption of Fig. \ref{fig: DeepSC_experiment_model_20230715}, $H$ is the number of heads of a Transformer encoder/decoder, and $V$ is the hidden layer dimension of the Transformer encoder's/decoder's feedforward network per Fig. \ref{fig: Trans_encoder_decoder_20230803}.  

Parameterized by $(K, H, V, E, L) = (8, 10, 32, 32, 30)$, the narrow DeepSC model, along with its layer parameters and connections, are schematized in \cite[Figs. 12 and 13]{Getu2023_empirical_study} (see \cite[Appendix A]{Getu2023_empirical_study}). Parameterized by $(K, H, V, E, L) = (8, 10, 32, 64, 30)$, the relatively wide DeepSC model, along with its layer parameters and connections, are diagrammed in \cite[Figs. 14 and 15]{Getu2023_empirical_study} (see \cite[Appendix A]{Getu2023_empirical_study}). In what follows, we explain how we define and generate the aforementioned two models using Keras with TensorFlow as a backend. 

We use the Keras model class \texttt{tf.keras.Model()} \cite{Keras_Model_class}\linebreak to define our two DeepSC models, which are returned by our DeepSC function that takes the target vocabulary size, $K$, $H$, $V$, $E$, and $L$ as its inputs. In this function, we first specify the input of a DeepSC model and its shape using the Keras function \texttt{tf.keras.Input()} \cite{Keras_Model_class}. This input layer is fed to an embedding layer, implemented using the Keras function \texttt{tf.keras.layers.Embedding()} \cite{Keras_Embedding_Layer}, with an input dimension $L$ and an output dimension (embedding dimension) $E$. The output of this embedding layer is fed to three cascaded Transformer encoders that we implement using the Keras function \texttt{keras$\_$nlp.layers.TransformerEncoder()} \cite{Keras_TransformerEncoder_Layer}. The three Transformer encoders are implemented to have $V$ and $H$ as the hidden size of their feedforward networks and the number of heads in their multi-head attention layers, respectively; the \texttt{he$\_$normal} initializer \cite{Keras_weight_initializers} as their kernel initializer; and a linear activation function. 

The three cascaded Transformer encoders' outputs are fed to two cascaded \texttt{Dense} layers that are realized using the Keras function \texttt{tf.keras.layers.Dense()} \cite{Keras_Dense_Layer}.\linebreak The two \texttt{Dense} layers have a ReLU activation function and the \texttt{he$\_$normal} initializer \cite{Keras_weight_initializers} as their kernel initializer. The output tensor of these cascaded \texttt{Dense} layers is reshaped into a matrix -- using the Keras reshape layer \texttt{tf.keras.layers.Reshape()} \cite{Keras_Reshape_Layer} -- that is fed to the non-trainable (linear) \texttt{Dense} layer, which models our presumed AWGN channel. Following \cite{O'Shea_TCCN'17}, we initialize this linear AWGN layer with the \texttt{Identity()} initializer \cite{Keras_weight_initializers} and \texttt{tf.keras.initializers.RandomNormal()} initializer \cite{Keras_weight_initializers} (with zero mean and variance 0.1) as its kernel initializer and bias initializer, respectively. The AWGN channel output is reshaped using the Keras reshape layer \texttt{tf.keras.layers.Reshape()} \cite{Keras_Reshape_Layer} before being fed to two other cascaded \texttt{Dense} layers realized using the Keras function \texttt{tf.keras.layers.Dense()} \cite{Keras_Dense_Layer}. These \texttt{Dense} layers also have a ReLU activation function and the \texttt{he$\_$normal} initializer \cite{Keras_weight_initializers} as their kernel initializer, realizing the channel decoder. 

The channel decoder's output is inputted to three cascaded Transformer decoders implemented using the function \texttt{keras$\_$nlp.layers.TransformerDecoder()} \cite{Keras_TransformerDecoder_Layer} that is set to have no cross-attention. The three Transformer decoders are also implemented to have $V$ and $H$ as the hidden layer size of their feedforward networks and the number of heads in their multi-head attention layers, respectively; the \texttt{he$\_$normal} initializer \cite{Keras_weight_initializers} as their corresponding kernel initializer; and a linear activation function. These cascaded Transformer decoders constitute a semantic decoder whose output is fed to a \texttt{softmax} prediction layer realized using the function \texttt{tf.keras.layers.Dense()} \cite{Keras_Dense_Layer}, with its output dimension equal to the target vocabulary size. 

Produced by cascading the above-detailed Keras functions, we generated narrow DeepSC using $(K, H, V, E, L)=(8, 10, 32, 32, 30)$ and relatively wide DeepSC using $(K, H, V, E, L)=(8, 10, 32, 64, 30)$. These models are \textit{compiled} using the \texttt{Adam} \cite{Keras_Adam_Opt} optimizer, \texttt{categorical$\_$crossentropy} loss, and accuracy as a classification metric, setting up our training detailed below.
\begin{table*}[!t]	
	\caption{DeepSC training (hyper)parameters unless otherwise mentioned.}		
	\label{table: Hyperparameters}
	\centering
	\begin{tabular}{ | l | l | l | } 
		\hline      
		\textbf{(Hyper)parameters} & \textbf{Type/Value} & \textbf{Remark(s)}     \\  \hline  
		Learning rate  & 0.0002    & This initial learning leads to our best DeepSC training performance.  \\  \hline      
		Epoch size &   100      & The size of the maximum epoch.      \\ \hline           
		Batch size & 50       & We try both large and small batch sizes. The latter yields a better training      \\    
		&    &   performance (though at a price of slow convergence).  \\      \hline 
		Optimizer & \texttt{Adam} \cite{Keras_Adam_Opt} & We experiment with \texttt{SGD with momentum} \cite{Keras_SGD_Opt}, \texttt{RMSprop} \cite{Keras_RMSprop_Opt}, and            \\   
		&    & \texttt{Adam} \cite{Keras_Adam_Opt}. \texttt{Adam} leads to our best training results.      \\   \hline
		Activation function & ReLU     & All \texttt{Dense} layers are equipped with a ReLU activation function except                    \\   
		(for \texttt{Dense} layers) &     &  the linear \texttt{Dense} layer that models the AWGN channel and the     \\   
		&     & \texttt{softmax} prediction layer, which is the last layer of DeepSC.                \\   \hline
		
		Layer weight initializer   & \texttt{he$\_$normal} \cite{Keras_weight_initializers} & Transformer encoder layers, Transformer decoder layers, and all \texttt{Dense}         \\   
		(for most layers) &     &  layers are initialized with \texttt{he$\_$normal} except the linear \texttt{Dense} layer            \\ 
		&    & that modeled our presumed AWGN channel. This layer was initialized       \\ 
		&    & with the Keras \texttt{Identity()} \cite{Keras_weight_initializers} initializer.   \\ \hline
		
		Bias initializer  & None    & No bias initializer is used on all DeepSC layers except the linear             \\   
		(for most layers) &     &  \texttt{Dense} layer -- which models our assumed AWGN channel -- initialized            \\   
		&    & by the Keras \texttt{RandomNormal()} initializer \cite{Keras_weight_initializers}.     \\   \hline
		
		$\sigma$    & 0.1    &  $\sigma^2=0.01$ W is the considered noise power during our training/testing.    \\   \hline
		
		Narrow DeepSC & $(K, H, V, E, L)=$     & The narrow DeepSC model, along with its layer parameters and       \\   
		&  $(8, 10, 32, 32, 30)$   & connections, are schematized in \cite[Figs. 12 and 13]{Getu2023_empirical_study}.           \\   \hline
		
		Relatively wide DeepSC & $(K, H, V, E, L)=$    & The relatively wide DeepSC model, along with its layer parameters        \\   
		& $(8, 10, 32, 64, 30)$    & and connections, are diagrammed in \cite[Figs. 14 and 15]{Getu2023_empirical_study}.             \\   \hline

		Training set size & 1.5 million    & The training set is prepared per the data tokenization, data vectorization,             \\   
		&     &  and data labeling of Secs. \ref{subsec: data_tokenization}, \ref{subsec: data_vectorization}, and \ref{subsec: data_labeling}, respectively.           \\   \hline

		Validation set size &  300,000    & The validation set is prepared per the data tokenization, data vectorization,        \\   
		&     & and data labeling of Secs. \ref{subsec: data_tokenization}, \ref{subsec: data_vectorization}, and \ref{subsec: data_labeling}, respectively.         \\   \hline

		Keras \textit{callbacks} \cite{Chollet_DL_with_Python'18} & Four callbacks  & We use \textit{Model checkpointing}, \textit{TensorBoard}, \textit{learning rate reduction}, and              \\   
		&     &   \textit{early stopping} callbacks \cite[Ch. 7]{Chollet_DL_with_Python'18}. Early stopping and learning rate          \\   
		&     & reduction are set to have patience over 10 and 5 epochs, respectively.     \\   
		&     & Per every five epochs that lead to a training stagnation w.r.t. the monitored   \\   
		&     & validation losses, the learning rate reduction Keras callback is set to     \\   
		&     & reduce the learning rate by 0.1.     \\   \hline
	\end{tabular}
\end{table*}
\subsection{Training Results of DeepSC}
\label{subsec: DeepSC_model_of_DeepSC}
Using the generated narrow DeepSC and relatively wide DeepSC models, we carry out extensive computer experiments on their training using \texttt{Adam} optimizer\footnote{We also perform extensive DeepSC training using \texttt{SGD with momentum} \cite{Keras_SGD_Opt} and \texttt{RMSprop} \cite{Keras_RMSprop_Opt}. Although we obtain much better training results with \texttt{SGD with Momentum} than with \texttt{RMSprop}, none of these optimizers leads to better training results than the ones we obtain with \texttt{Adam}.} \cite{Keras_Adam_Opt} and the (hyper)parameters in Table \ref{table: Hyperparameters}, such as the four \textit{Keras callbacks} \cite{Chollet_DL_with_Python'18}. Specifically, we deploy the Keras \texttt{model.fit$\_$generator()} \cite{Keras_Model_Training_APIs} function that is fed with the training and validation sets (generated per Secs. \ref{subsec: data_tokenization}, \ref{subsec: data_vectorization}, and \ref{subsec: data_labeling}), maximum epochs and steps per epoch, validation steps, and Keras callbacks to extensively train both the narrow DeepSC and relatively wide DeepSC models.
	
Our final training is conducted in Graham using 4 GPUs simultaneously, in line with a distributed training strategy called \texttt{with tf.distribute.MirroredStrategy()}. Particularly, we deploy Graham's four NVIDIA T4 Turing GPUs (16GB memory) \cite{DRAC_Graham} in parallel to train both the narrow DeepSC model (see \cite[Fig. 12]{Getu2023_empirical_study}) and the relatively wide DeepSC model (see \cite[Fig. 14]{Getu2023_empirical_study}) for about four days. This extensive training and optimization led to narrow DeepSC's and relatively wide DeepSC's training results reported in Secs. \ref{subsubsec: Training_results_of_narrow_DeepSC} and \ref{subsubsec: Training_results_of_relatively_wide_DeepSC}, respectively.	  
\begin{figure*}[t!]
	\begin{minipage}[htb]{0.45\linewidth}
		\centering
		\includegraphics[width=\textwidth]{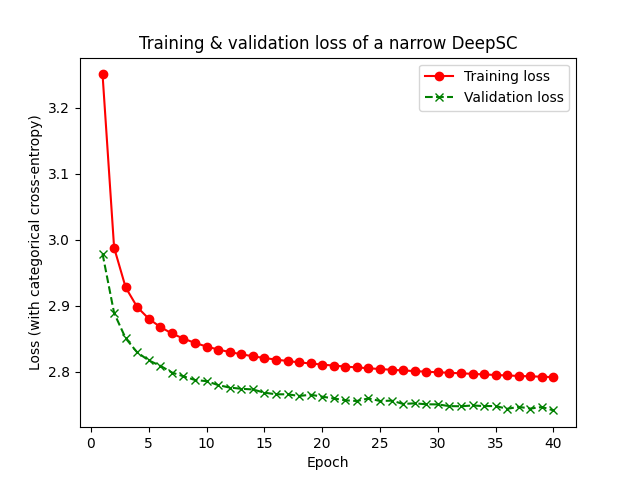}   \\ [-0.25cm]
		\caption{Training and validation loss of narrow DeepSC when the considered maximum epoch size is 40.} 
		\label{fig: Training-0815-I2_loss_of_DeepSC}
	\end{minipage}
	\hspace{0.5cm}
	\begin{minipage}[htb]{0.45\linewidth}
		\centering
		\includegraphics[width=\textwidth]{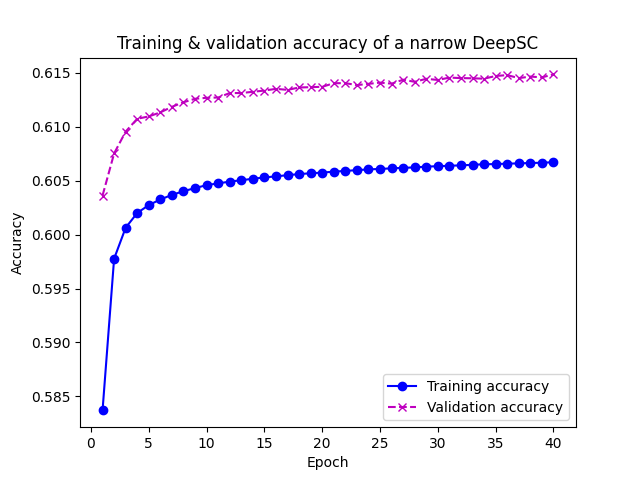}   \\ [-0.25cm]
		\caption{Training and validation accuracy of narrow DeepSC when the considered maximum epoch size is 40.} 
		\label{fig: Training-0815-I2_accuracy_of_DeepSC}	
	\end{minipage}	
\end{figure*} 

\begin{figure*}[t!]
	\begin{minipage}[htb]{0.45\linewidth}
		\centering
		\includegraphics[width=\textwidth]{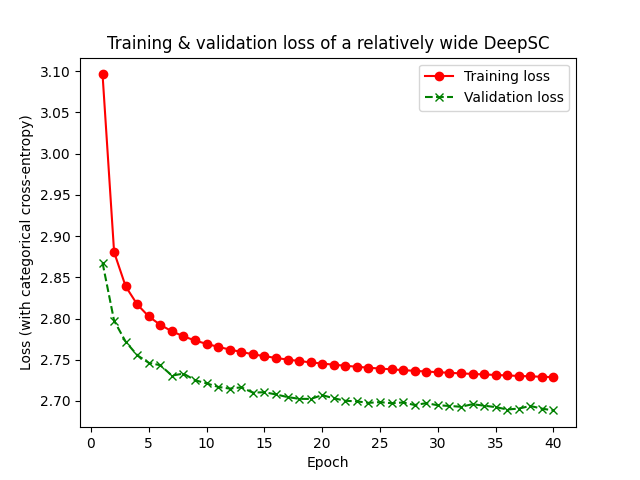}   \\ [-0.25cm]
		\caption{Training and validation loss of relatively wide DeepSC when the considered maximum epoch size is 40.} 
		\label{fig: Training-0815-I1_loss_of_DeepSC}
	\end{minipage}
	\hspace{0.5cm}
	\begin{minipage}[htb!]{0.45\linewidth}
		\centering
		\includegraphics[width=\textwidth]{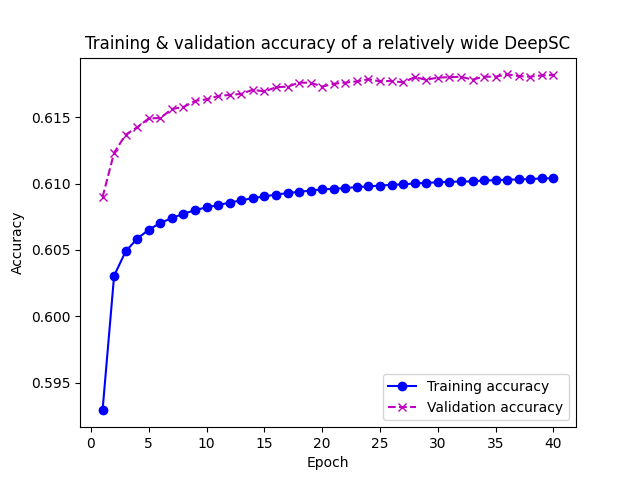}   \\ [-0.25cm]
		\caption{Training and validation accuracy of relatively wide DeepSC when the considered maximum epoch size is 40.} 
		\label{fig: Training-0815-I1_accuracy_of_DeepSC}	
	\end{minipage}	
\end{figure*}

\subsubsection{Training Results of Narrow DeepSC}
\label{subsubsec: Training_results_of_narrow_DeepSC}
Figs. \ref{fig: Training-0815-I2_loss_of_DeepSC} and \ref{fig: Training-0815-I2_accuracy_of_DeepSC} show the training and validation loss and the training and validation accuracy of the narrow DeepSC model, respectively. These results are the best ones that we achieve for narrow DeepSC with the (hyper)parameters in Table \ref{table: Hyperparameters}, especially, with 1.5 million and 300,000 training and validation sentences, respectively. As can be seen in Fig. \ref{fig: Training-0815-I2_loss_of_DeepSC}, the training and validation loss of narrow DeepSC does not meaningfully decrease after 20 epochs, where there is a very small improvement for every epoch exceeding 20.

Fig. \ref{fig: Training-0815-I2_accuracy_of_DeepSC} shows that the training and validation accuracy of narrow DeepSC consistently improves until the end of the 40-th epoch, especially the validation accuracy. By the 40-th epoch, narrow DeepSC attains a training and validation accuracy slightly higher than 60.5\% and nearly 61.5\%, respectively. Although 61.5\% is hardly a high validation accuracy, the result is considerable for a DL-based multi-class classification problem with 22,899 classes of words, corresponding to the dictionary size per \cite[Appendix A]{Getu2023_empirical_study}.
\begin{figure*}[t!]
	\begin{minipage}[htb]{0.45\linewidth}
		\centering
		\includegraphics[width=\textwidth]{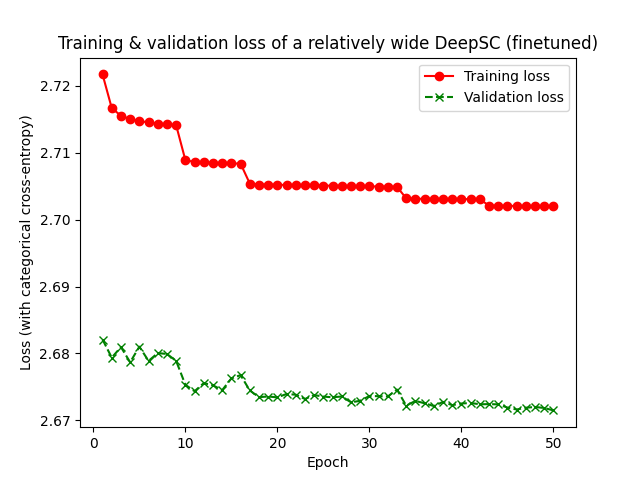}   \\ [-0.25cm]
		\caption{Training and validation loss of relatively wide trained and finetuned DeepSC when the considered initial learning rate and maximum epoch size are 0.0001 and 50, respectively.} 
		\label{fig: Training-0822-I1_loss_of_DeepSC}
	\end{minipage}
	\hspace{0.5cm}
	\begin{minipage}[htb]{0.45\linewidth}
		\centering
		\includegraphics[width=\textwidth]{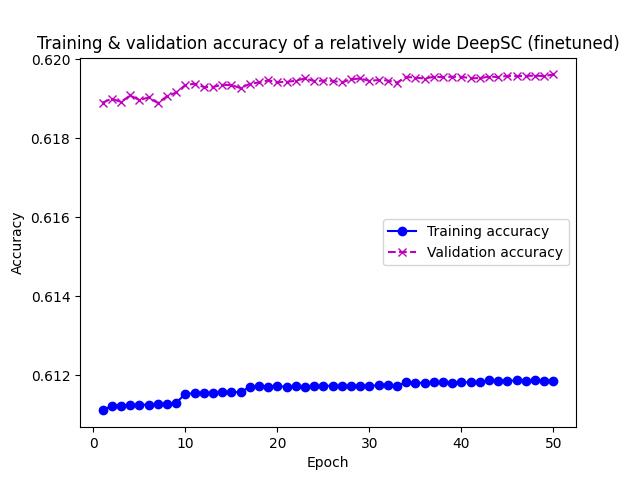}   \\ [-0.25cm]
		\caption{Training and validation accuracy of relatively wide trained and finetuned DeepSC when the considered initial learning rate and maximum epoch size are 0.0001 and 50, respectively.} 
		\label{fig: Training-0822-I1_accuracy_of_DeepSC}	
	\end{minipage}	
\end{figure*}

\subsubsection{Training Results of Relatively Wide DeepSC}
\label{subsubsec: Training_results_of_relatively_wide_DeepSC}
Figs. \ref{fig: Training-0815-I1_loss_of_DeepSC} and \ref{fig: Training-0815-I1_accuracy_of_DeepSC} show the training and validation loss and the training and validation accuracy of the relatively wide DeepSC model, respectively. These results are the best ones that we obtain for relatively wide DeepSC with the (hyper)parameters in Table \ref{table: Hyperparameters}, especially, with 1.5 million and 300,000 training and validation sentences, respectively. As shown in Fig. \ref{fig: Training-0815-I1_loss_of_DeepSC}, relatively wide DeepSC manifests a training stagnation after the 25-th epoch w.r.t. its training and validation loss, despite a small improvement for every epoch exceeding 25.

Fig. \ref{fig: Training-0815-I1_accuracy_of_DeepSC} shows that the training and validation accuracy of relatively wide DeepSC also steadily improves until the 40-th epoch, especially the validation accuracy. By the end of the 40-th epoch, relatively wide DeepSC achieves a training and validation accuracy of nearly 61\% and 62\%, respectively. These results demonstrate an improvement as compared to the training performance of narrow DeepSC, and are thus crucial for a DL-based multi-class classification problem with 22,899 classes (of words). Seeking to produce trained DeepSC models that would achieve better classification accuracy, we also train both narrow DeepSC and relatively wide DeepSC models -- also using Table \ref{table: Hyperparameters}'s parameters -- for seven days using Graham's four NVIDIA T4 Turing (16GB memory) GPUs \cite{DRAC_Graham} in parallel. These narrow DeepSC and relatively wide DeepSC trained models slightly improve narrow DeepSC's and relatively wide DeepSC's training/validation performance shown in Figs. \ref{fig: Training-0815-I2_loss_of_DeepSC}-\ref{fig: Training-0815-I2_accuracy_of_DeepSC}\linebreak and Figs. \ref{fig: Training-0815-I1_loss_of_DeepSC}-\ref{fig: Training-0815-I1_accuracy_of_DeepSC}, respectively. Similarly, the seven-day-trained relatively wide DeepSC model produces better training and validation accuracy than the seven-day-trained narrow DeepSC. We refer to this relatively wide DeepSC model \textquotedblleft DeepSC-Training-0814-I1'', which we finetune as follows.

\subsubsection{Training Results of Relatively Wide Trained and Finetuned DeepSC}
After setting all layers (except for the AWGN channel layer) of the trained relatively wide DeepSC model named \textquotedblleft DeepSC-Training-0814-I1'' to trainable mode, we finetune \textquotedblleft DeepSC-Training-0814-I1'' for nearly five days using Graham's four NVIDIA T4 Turing (16GB memory) GPUs \cite{DRAC_Graham} in parallel. This led to Figs. \ref{fig: Training-0822-I1_loss_of_DeepSC} and \ref{fig: Training-0822-I1_accuracy_of_DeepSC} that show the training and validation loss of the relatively wide trained and finetuned DeepSC and the training and validation accuracy of relatively wide trained and finetuned DeepSC, respectively. Concerning the former, Fig. \ref{fig: Training-0822-I1_loss_of_DeepSC} shows that the relatively wide trained and finetuned DeepSC delivers a 0.02 and 0.01 improvement in the training and validation loss, respectively, at the end of the 50-th\linebreak epoch (after nearly five days of training). By the same token, Fig. \ref{fig: Training-0822-I1_accuracy_of_DeepSC} demonstrates that the relatively wide trained and finetuned DeepSC produces nearly a 0.01 improvement in training and validation accuracy, after nearly five days of training.

Meanwhile, we name our relatively wide trained and finetuned DeepSC model \textquotedblleft DeepSC-Training-0822-I1'' which we also employ in our testing along with our already trained relatively wide DeepSC model named \textquotedblleft DeepSC-Training-0814-I1'', as reported below. 

\section{Testing of the Trained DeepSC Models with(out) Multi-Interferer RFI}
\label{sec: DeepSC_testing} 
This section details our testing setup and assumptions, testing procedures, and testing results, beginning with our testing setup and assumptions. 
\subsection{Testing Setup and Assumptions} 
\label{subsec: testing_setup_and_assumptions}
The testing of the trained relatively wide DeepSC models introduced above -- i.e., \textquotedblleft DeepSC-Training-0814-I1'' and \textquotedblleft DeepSC-Training-0822-I1'' -- considers the cases where no RFI and MI RFI (from $U$ RFI emitters) are received by the DeepSC receiver. For these models, the respective DeepSC symbols are equated as    % 
\begin{subequations}
\begin{align}
\label{x_tilde_vec_defn}
\tilde{\bm{x}}_t  & \eqdef C_{\hat{\bm{\alpha}}} (S_{\hat{\bm{\beta}}}(\tilde{\bm{s}}_t))   \in \mathbb{R}^{1 \times KL}       \\ 
\label{x_tilde_f_vec_defn}
\tilde{\bm{x}}_t^f  & \eqdef C_{\tilde{\bm{\alpha}}}^f  (S_{\tilde{\bm{\beta}}}^f (\tilde{\bm{s}}_t))   \in \mathbb{R}^{1 \times KL}, 
\end{align}
\end{subequations}
where $\tilde{\bm{x}}_t$ and $\tilde{\bm{x}}_t^f$ are the DeepSC symbols of the relatively wide trained DeepSC model and the relatively wide trained and finetuned DeepSC model, respectively; $S_{\hat{\bm{\beta}}}(\cdot)$ and $C_{\hat{\bm{\alpha}}}(\cdot)$ are the relatively wide trained DeepSC model's semantic encoder network with a parameter set $\hat{\bm{\beta}}$ and channel encoder network with a parameter set $\hat{\bm{\alpha}}$, respectively; $S_{\tilde{\bm{\beta}}}^f(\cdot)$ and $C_{\tilde{\bm{\alpha}}}^f(\cdot)$ are the relatively wide trained and finetuned DeepSC model's semantic encoder network with a parameter set $\tilde{\bm{\beta}}$ and channel encoder network with a parameter set $\tilde{\bm{\alpha}}$, respectively; and $\tilde{\bm{s}}_t$ is the $t$-th testing sentence encoded and prepared according to Sec. \ref{subsec: preparation_of_datasets}. For the case of no MI RFI, the respective received DeepSC signals during the $t$-th time slot are given by
\begin{subequations}
\begin{align}
\label{y_tilde_vec_defn}
\tilde{\bm{y}}_t   &  \eqdef \tilde{\bm{x}}_t + \bm{n}  \in \mathbb{R}^{1 \times KL}    \\ 
\label{y_tilde_f_vec_defn}
\tilde{\bm{y}}_t^f & \eqdef \tilde{\bm{x}}_t^f + \bm{n}  \in \mathbb{R}^{1 \times KL},
\end{align}
\end{subequations}
where $\tilde{\bm{y}}_t$ and $\tilde{\bm{y}}_t^f$ are the DeepSC signals of the relatively wide trained DeepSC model and the relatively wide trained and finetuned DeepSC model, respectively.
 
For the MI RFI scenario, we presume -- without loss of generality -- that the trained channel decoders would receive time-varying MI RFI from $U$ Gaussian RFI (i.e., broadband RFI) \cite{GeAR_conf_16,TMGWA17,TMWAR_TWC_18,TWR_WCL_2018} emitters whose interference signals are received over Rayleigh fading channels. Under MI RFI, the received DeepSC signals during the $t$-th time slot are equated as  
\begin{subequations}
\begin{align}
\label{y_tilde_vec_defn_with_MI-RFI}
\tilde{\bm{y}}_t   &  \eqdef \tilde{\bm{x}}_t + \sum_{u=1}^U g_{u,t} \bm{v}_{u,t} + \bm{n}   \in \mathbb{R}^{1 \times KL}       \\ 
\label{y_tilde_f_vec_defn_with_MI-RFI}
\tilde{\bm{y}}_t^f   &  \eqdef \tilde{\bm{x}}_t^f + \sum_{u=1}^U g_{u,t}^f \bm{v}_{u,t}^f + \bm{n}   \in \mathbb{R}^{1 \times KL},
\end{align}
\end{subequations}
where $\tilde{\bm{x}}_t$ and $\tilde{\bm{x}}_t^f$ are defined in (\ref{x_tilde_vec_defn}) and (\ref{x_tilde_f_vec_defn}), respectively; $g_{u,t}, g_{u,t}^f \sim \mathcal{N} (0,1)$ are the channel coefficients\footnote{During the reception of the $t$-th DeepSC symbol, the channel coefficients $g_{u,t}, g_{u,t}^f \sim \mathcal{N}(0,1)$ remain constant for the duration of the $t$-th DeepSC symbol, which is equal to $KL$ times the duration of each semantic symbol, signifying slowly-varying MI RFI.} of the $u$-th RFI during the $t$-th time slot; and $(\bm{v}_{u,t})_i, (\bm{v}_{u,t}^f)_i \sim \mathcal{N} (0, 10)$ are the respective Gaussian RFI signals -- with an assumed power that is equal to 10 W -- of the $u$-th Gaussian RFI emitted during the $t$-th time slot $\forall i \in [KL]$. As defined in (\ref{y_tilde_vec_defn_with_MI-RFI}) and (\ref{y_tilde_f_vec_defn_with_MI-RFI}), $\tilde{\bm{y}}_t$ and $\tilde{\bm{y}}_t^f$ go through -- as in Fig. \ref{fig: DeepSC_experiment_model_20230715} -- the trained channel decoders whose outputs are fed to the trained semantic decoders to produce the $t$-th recovered testing sentences $\hat{\tilde{\bm{s}}}_t$ and $\hat{\tilde{\bm{s}}}_t^f$ that can be expressed as 
\begin{subequations}
\begin{align}
\label{s_tilde_hat_vec_defn}
\hat{\tilde{\bm{s}}}_t  &  \eqdef S_{\hat{\bm{\theta}}} (C_{\hat{\bm{\delta}}}(\tilde{\bm{y}}_t))   \\
\label{s_tilde_hat_f_vec_defn}
\hat{\tilde{\bm{s}}}_t^f  &  \eqdef S_{\tilde{\bm{\theta}}}^f (C_{\tilde{\bm{\delta}}}^f(\tilde{\bm{y}}_t^f)),
\end{align}	
\end{subequations}
where $C_{\hat{\bm{\delta}}} (\cdot)$ and $S_{\hat{\bm{\theta}}} (\cdot)$ are the relatively wide trained DeepSC model's channel decoder network with a parameter set $\hat{\bm{\delta}}$ and semantic decoder network with a parameter set $\hat{\bm{\theta}}$, respectively; $C_{\tilde{\bm{\delta}}}^f (\cdot)$ and $S_{\tilde{\bm{\theta}}}^f (\cdot)$ are the relatively wide trained and finetuned DeepSC model's channel decoder network with a parameter set $\tilde{\bm{\delta}}$ and semantic decoder network with a parameter set $\tilde{\bm{\theta}}$, respectively. Now, $\{ ( \tilde{\bm{s}}_t, \hat{\tilde{\bm{s}}}_t ) \}$ and $\{ ( \tilde{\bm{s}}_t, \hat{\tilde{\bm{s}}}_t^f ) \}$ have to be compared from a semantic vantage point $\forall t \in \mathbb{N}$.

Although our training is conducted using categorical cross-entropy loss while setting classification accuracy as a metric, this performance assessment metric is not a proper semantic metric for a text SemCom \cite{Getu_IEEE_Access'23}. Consequently, we adopt our recently proposed semantic metric, named the \textit{(upper tail) probability of semantic similarity} $p(\eta_{\textnormal{min}})$ \cite{Getu_TWC'24}, to evaluate the testing performance of the relatively wide trained DeepSC model and the relatively wide trained and finetuned DeepSC model w.r.t. a minimum semantic similarity $\eta_{\textnormal{min}} \in [0,1]$. This metric is defined for the $t$-th transmitted and recovered sentence pairs $\{ ( \tilde{\bm{s}}_t, \hat{\tilde{\bm{s}}}_t ) \}$ and $\{ ( \tilde{\bm{s}}_t, \hat{\tilde{\bm{s}}}_t^f ) \}$ as \cite[eq. (9)]{Getu_TWC'24}  
\begin{subequations}
\begin{align}
\label{PSS_def}
p(\eta_{\textnormal{min}})   &  \eqdef \mathbb{P}(\eta(\tilde{\bm{s}}_t, \hat{\tilde{\bm{s}}}_t) \geq \eta_{\textnormal{min}} )  \\ 
\label{PSS_F_def}
p(\eta_{\textnormal{min}})   &  \eqdef \mathbb{P}(\eta(\tilde{\bm{s}}_t, \hat{\tilde{\bm{s}}}_t^f ) \geq \eta_{\textnormal{min}}), 
\end{align}
\end{subequations}
where $\eta(\tilde{\bm{s}}_t, \hat{\tilde{\bm{s}}}_t)$ denotes the semantic similarity between $ \tilde{\bm{s}}_t$ and $\hat{\tilde{\bm{s}}}_t $; $\eta(\tilde{\bm{s}}_t, \hat{\tilde{\bm{s}}}_t^f )$ stands for the semantic similarity between $ \tilde{\bm{s}}_t$ and $\hat{\tilde{\bm{s}}}_t^f$; and $\eta(\cdot, \cdot)$ is the semantic similarity (or sentence similarity) function that is often estimated using large pre-trained Transformers, such as the \textit{HuggingFace}'s (see \cite{HuggingFace_Company}) \textit{sentence Transformers} \cite{HuggingFace_Sen_Transformers}. Among these Transformers, we use -- without loss of generality -- HuggingFace's lightweight state-of-the-art sentence Transformer dubbed \texttt{all-MiniLM-L6-v2} \cite{HuggingFace_all-MiniLM-L6-v2}.\footnote{\texttt{all-MiniLM-L6-v2} is an important sentence Transformer applicable to semantic similarity estimation and clustering or semantic search \cite{HuggingFace_all-MiniLM-L6-v2}. It works by mapping paragraphs and sentences to a 384-dimensional dense vector space \cite{HuggingFace_all-MiniLM-L6-v2}.}

Using our testing sentences (encoded and prepared per\linebreak Sec. \ref{subsec: preparation_of_datasets}), we numerically estimate -- w.r.t. (\ref{PSS_def}) and (\ref{PSS_F_def}) -- the probability of semantic similarity exhibited by the relatively wide trained DeepSC model and the relatively wide trained and finetuned DeepSC model as 
\begin{subequations}
	\begin{align}
		\label{PSS_def_2}
		p(\eta_{\textnormal{min}})   &  = \frac{1}{N} \sum_{t=1}^N \mathbb{I}\{ \eta(\tilde{\bm{s}}_t, \hat{\tilde{\bm{s}}}_t) \geq \eta_{\textnormal{min}} \}     \\ 
		\label{PSS_F_def_2}
		p(\eta_{\textnormal{min}})   &  = \frac{1}{N} \sum_{t=1}^N \mathbb{I}\{ \eta(\tilde{\bm{s}}_t, \hat{\tilde{\bm{s}}}_t^f ) \geq \eta_{\textnormal{min}} \}, 
	\end{align}
\end{subequations}
where $N$ stands for the number of testing sentences used to assess the manifested probability of semantic similarity. Concerning (\ref{PSS_def_2}) and (\ref{PSS_F_def_2}), we compute $\eta(\tilde{\bm{s}}_t, \hat{\tilde{\bm{s}}}_t)$ and $\eta(\tilde{\bm{s}}_t, \hat{\tilde{\bm{s}}}_t^f )$ using the outputs of the sentence Transformer \texttt{all-MiniLM-L6-v2} -- denoted by $\bm{T}_{\bm{\Phi}}(\cdot)$ -- as 
\begin{subequations}  %  
\begin{align}
\label{Cosine_similarity_1}
\eta(\tilde{\bm{s}}_t, \hat{\tilde{\bm{s}}}_t) & =  \frac{\bm{T}_{\bm{\Phi}}(\tilde{\bm{s}}_t) \cdot  \bm{T}_{\bm{\Phi}}(\hat{\tilde{\bm{s}}}_t)^T }{ \| \bm{T}_{\bm{\Phi}}(\tilde{\bm{s}}_t)\|  \| \bm{T}_{\bm{\Phi}}(\hat{\tilde{\bm{s}}}_t) \| }	    \\ 
\label{Cosine_similarity_f_1}
\eta(\tilde{\bm{s}}_t, \hat{\tilde{\bm{s}}}_t^f ) & = \frac{ \bm{T}_{\bm{\Phi}}(\tilde{\bm{s}}_t) \cdot \bm{T}_{\bm{\Phi}}(\hat{\tilde{\bm{s}}}_t^f)^T }{\| \bm{T}_{\bm{\Phi}}(\tilde{\bm{s}}_t)\|  \| \bm{T}_{\bm{\Phi}}(\hat{\tilde{\bm{s}}}_t^f) \|}, 
\end{align}
\end{subequations} 
where (\ref{Cosine_similarity_1}) and (\ref{Cosine_similarity_f_1}) compute the \textit{cosine similarity} w.r.t. the adopted sentence Transformer's output.

In light of the above-detailed testing setup and assumptions, the prepared testing sentences of Sec. \ref{subsec: preparation_of_datasets}, and the tokenizer of Sec. \ref{subsec: data_tokenization}, we move on to detail our testing procedures.    

\subsection{Testing Procedures}
\label{subsec: Testing_procedures}
This subsection systematically details our testing procedures such as mapping an integer to a word, prediction of a recovered sentence, and evaluation of the trained DeepSC models using the probability of semantic similarity computed numerically in (\ref{PSS_def_2}) and (\ref{PSS_F_def_2}) via (\ref{Cosine_similarity_1}) and (\ref{Cosine_similarity_f_1}). We begin with our function that maps an integer to a word.
\subsubsection{Mapping an Integer to a Word}
\label{subsubsec: mapping_an_integer}
After attempting to learn the probability of each word of a sentence (via \texttt{softmax} layers), the relatively wide trained DeepSC model and the relatively wide trained and finetuned DeepSC model return the most likely integer for every word. As every integer except zero encodes a unique word per our employed Keras tokenizer, we implement a Python function that returns a word for a predicted integer. This function accepts an integer index and the tokenizer of Sec. \ref{subsec: data_tokenization}, and returns a word when a word's index matches the inputted integer index. Otherwise, this function returns nothing.

Appending and joining each word predicted by a trained model produces a recovered sentence, as explained below.  
\subsubsection{Prediction of a Recovered Sentence}
\label{subsubsec: recovered_sentence_prediction}
To determine the recovered sentence of a trained DeepSC model for a given input sentence, the trained network should first compute the probability for each word of the input sentence. Based on this probability, we can employ our Keras tokenizer -- used to prepare our training, validation, and testing sets (i.e., the tokenizer of Sec. \ref{subsec: data_tokenization}) -- to determine the likely word to each tokenized and transmitted word. Joining and appending each likely word, one can infer the recovered sentence. Deploying this idea to generate a recovered sequence given the transmitted sequence (source sequence) of words, we write a Python function that takes a trained DeepSC model, the tokenizer of Sec. \ref{subsec: data_tokenization}, and a source sequence of integers (\textquotedblleft source'') as its inputs. This function returns the respective recovered sequence by implementing the following procedures:
\begin{enumerate}
	\item This function first generates a prediction matrix of size $L \times 22,899$ using the Keras function and code \texttt{model.predict(source)[0]}. 
	
	\item It then generates the predicted sequence of integers by applying the function \texttt{argmax($\cdot$)} to every column vector of the already obtained prediction matrix.
	
	\item For every predicted integer in a \texttt{for loop}, it then generates a word using the function described in Sec. \ref{subsubsec: mapping_an_integer}.\linebreak If the returned word is none, the \texttt{for loop} breaks; if not, the \texttt{for loop} continues to append each generated word until the last integer's predicted word is appended and the \texttt{for loop} is ended. Upon ending the \texttt{for loop}, the function joins the appended words and returns the recovered sentence.
\end{enumerate}

The mentioned steps return a recovered sentence for every transmitted testing sentence. However, we have 100,000 testing sentences (prepared per Sec. \ref{subsec: preparation_of_datasets}), and the evaluation of our trained DeepSC models requires \textit{model update} per every testing time slot in case of MI RFI. Model update per each time slot is needed since our considered MI RFI from $U$ Gaussian emitters -- according to (\ref{y_tilde_vec_defn_with_MI-RFI}) and (\ref{y_tilde_f_vec_defn_with_MI-RFI}) -- varies in each time slot. Such time-varying MI RFI has to be accommodated during testing, as detailed below.

\subsubsection{Evaluation of the Trained DeepSC Models}
\label{subsubsec: Trained_DeepSC_model_evaluation}
Evaluation of our relatively wide trained DeepSC model and our relatively wide trained and finetuned DeepSC model is carried out using a Python function that we write to assess the performance of a trained DeepSC model. This Python function takes a trained DeepSC model, the tokenizer of Sec. \ref{subsec: data_tokenization}, the testing sequences of Sec. \ref{subsec: preparation_of_datasets}, the corresponding raw Europarl testing sentences, different values of $U$, $K$, and $L$, and returns its computed probability of semantic similarity exhibited by an inputted trained DeepSC model. Because this function evaluates the trained model -- with and without the time-varying MI RFI -- that is fed to it, its first step is model updating per the possible reception of time-varying MI RFI during testing, as detailed below.

If $U=0$ (no RFI), the trained DeepSC model evaluation function takes the inputted trained model as an updated model. If $U>0$, the same function handles the reception of time-varying MI RFI via an MI RFI linear \texttt{Dense} layer that is inserted -- per every testing time slot $t$ -- to the already trained DeepSC models that are fed to our model evaluation function. This is accomplished using \textit{three consecutive steps}\footnote{To accommodate the time-varying MI RFI per (\ref{y_tilde_vec_defn_with_MI-RFI}) and (\ref{y_tilde_f_vec_defn_with_MI-RFI}), we execute trained model disassembling; programming and insertion of an MI RFI linear \texttt{Dense} layer; and trained model reassembling once per every time slot $t$.} that are executed once per every testing time slot $t$: $i)$ Trained model disassembling; $ii)$ programming and insertion of an MI RFI linear \texttt{Dense} layer; and $iii)$ trained model reassembling. Trained model disassembling is accomplished using the following Python code snippet:  \\
\indent \texttt{layers = [l for l in model.layers]}. \\ 
To program an MI RFI linear \texttt{Dense} layer with the magnitude of the $t$-th MI RFI that is received w.r.t. a given $(U, K, L)$ tuple, we first execute the following Python function that computes the $t$-th total Gaussian MI RFI received over Rayleigh fading channels. \\
\indent \texttt{def Total$\_$RFI(U, K, L):} \\
\indent \indent \texttt{RFI=np.zeros([K*L])}  \\
\indent \indent \texttt{for j in range(U):}  \\
\indent \indent \indent \texttt{ RFI=RFI+np.random.normal(0, 1, 1)*\linebreak math.sqrt(10)*np.random.normal(0, 1, K*L)}  \\
\indent \indent \texttt{MI$\_$RFI=RFI}   \\
\indent \indent \texttt{return MI$\_$RFI}   \\
We then define the $t$-th \textit{non-trainable} MI RFI linear \texttt{Dense} layer (using the Keras \texttt{Dense} layer \cite{Keras_Dense_Layer}) by employing an identity weight matrix and biases that are equal to the $t$-th total Gaussian MI RFI vector's computed values. We then insert this layer into the first \texttt{Reshape} layer's output of the disassembled trained DeepSC architecture (per Fig. \ref{fig: DeepSC_experiment_model_20230715}) as: \\
\indent \texttt{MI$\_$RFI$\_$Layer=Dense(K*L, weights = [np.identity(K*L), Total$\_$RFI(U, K, L)], activation='linear', trainable=False, name='MI-RFI$\_$Linear$\_$Dense$\_$Layer')\linebreak (layers[5].output)}. \\
We then reassemble the inputted trained model with the inserted $t$-th MI RFI linear \texttt{Dense} layer, while setting all the constituting layers to be non-trainable:  \\
\indent \texttt{x = MI$\_$RFI$\_$Layer}  \\
\indent \texttt{for i in range(len(layers)):} \\
\indent \indent  \texttt{layers[i].trainable = False}  \\ 
\indent \indent \texttt{if i>5:}  \\
\indent \indent \indent \texttt{x = layers[i](x)}.  \\
At last, we obtain an updated model -- for the $t$-th testing time slot -- via the following code snippet:   \\   
\indent \texttt{Updated$\_$model = Model(inputs=layers[0].\linebreak input, outputs=x)}.  

Once an updated model for the $t$-th testing time slot is obtained, the trained DeepSC model evaluation function executes the following steps -- for every $t$-th sequence of Sec. \ref{subsec: preparation_of_datasets}'s testing sequences -- to compute the exhibited probability of semantic similarity by a given trained DeepSC model:
\begin{enumerate}
	\item  First, it generates the $t$-th recovered sentence using the sentence prediction function of Sec. \ref{subsubsec: recovered_sentence_prediction}.
	
	\item  Second, it feeds the $t$-th recovered sentence and the $t$-th (truncated) raw Europarl testing sentence -- truncated to length $L=30$ if $L>30$ -- to the sentence Transformer \texttt{all-MiniLM-L6-v2} \cite{HuggingFace_all-MiniLM-L6-v2}.
	
	\item  Third, it then feeds the $t$-th two embedded sentences of \texttt{all-MiniLM-L6-v2} to a cosine similarity utility function to compute the $t$-th semantic similarity value according to (\ref{Cosine_similarity_1}) or (\ref{Cosine_similarity_f_1}).
	
\end{enumerate}
Looping over all the aforementioned routines of Sec. \ref{subsubsec: Trained_DeepSC_model_evaluation} for $N$ testing time slots, our trained DeepSC model evaluation function would determine the manifested probability of semantic similarity -- by a trained DeepSC model -- according to (\ref{PSS_def_2}) or (\ref{PSS_F_def_2}). 

By inserting the relatively wide trained DeepSC model, the tokenizer of Sec. \ref{subsec: data_tokenization}, the testing sequences of Sec. \ref{subsec: preparation_of_datasets},\linebreak the corresponding raw Europarl testing sentences, different values of $U$, $K=8$, and $L=30$ into the DeepSC model evaluation function of Sec. \ref{subsubsec: Trained_DeepSC_model_evaluation}, we generate the $t$-th relatively wide trained DeepSC model with the $t$-th MI RFI linear \texttt{Dense} layer in \cite[Fig. 16]{Getu2023_empirical_study} (\cite[Appendix A]{Getu2023_empirical_study}) -- plotted for $U=50$ -- and the testing results of Sec. \ref{subsubsec: Testing_results_with_trained_DeepSC}.\linebreak In addition, we obtain the $t$-th relatively wide trained and finetuned DeepSC model with the $t$-th MI RFI linear \texttt{Dense} layer in \cite[Fig. 17]{Getu2023_empirical_study} (\cite[Appendix A]{Getu2023_empirical_study}) -- plotted for $U=50$ -- and the results reported in Sec. \ref{subsubsec: Testing_results_with_trained_and_finetuned_DeepSC} by inputting the relatively wide trained and finetuned DeepSC model, the tokenizer of Sec. \ref{subsec: data_tokenization}, the testing sequences of Sec. \ref{subsec: preparation_of_datasets}, the respective raw Europarl testing sentences, different values of $U$, $K=8$, and $L=30$ into the DeepSC model evaluation function of Sec. \ref{subsubsec: Trained_DeepSC_model_evaluation}.

\begin{figure*}[t!]
	\begin{minipage}[htb]{0.45\linewidth}
		\centering
		\includegraphics[width=\textwidth]{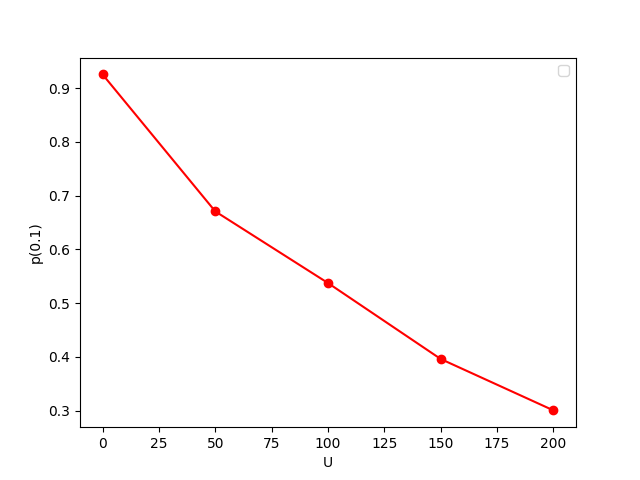}   \\ [-0.25cm]
		\caption{$p(0.1)$ versus $U$ manifested by our relatively wide trained DeepSC model for $N=10,000$ testing sentences.} 
		\label{fig: Prob_SSM-10000-0906-I}
	\end{minipage}
	\hspace{0.5cm}
	\begin{minipage}[htb]{0.45\linewidth}
		\centering
		\includegraphics[width=\textwidth]{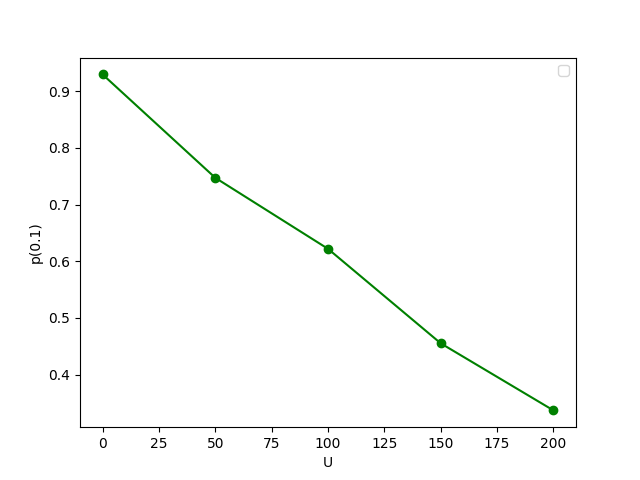}   \\ [-0.25cm]
		\caption{$p(0.1)$ versus $U$ manifested by our relatively wide trained and finetuned DeepSC model for $N=10,000$ testing sentences.} 
		\label{fig: Prob_SSM-10000-0906-II}	
	\end{minipage}	
\end{figure*}
  
\subsection{Testing Results of the Trained DeepSC Models with and without MI RFI}
\label{subsec: Testing_results}
During our extensive testing, we witnessed many recovered sentences that are surely unrelated with their transmitted counterparts, especially for large $U$. However, when these semantically unrelated sentences are fed to our adopted sentence Transformer, the resulting sentence embeddings lead to a semantic similarity of around 0.1 (when the embeddings are fed to the cosine similarity utility function). To take this limitation in our probabilistic assessment of semantic irrelevance, we plot $p(0.1)$ versus $U$ rather than $p(0)$ versus $U$, as documented below.
 
\subsubsection{Testing Results of the Relatively Wide Trained DeepSC Model} 
\label{subsubsec: Testing_results_with_trained_DeepSC}
Fig. \ref{fig: Prob_SSM-10000-0906-I} depicts the $p(0.1)$ versus $U$ plot manifested by our relatively wide trained DeepSC model tested using 10,000 testing sentences. As can be seen in Fig. \ref{fig: Prob_SSM-10000-0906-I}, our relatively wide trained DeepSC model produces semantically irrelevant sentences as the number of MI RFI interferers becomes large. This trend is consistent with the trend predicted by our recently developed theory \cite{Getu_TWC'24}.
 
\subsubsection{Testing Results of the Relatively Wide Trained and Finetuned DeepSC Model}
\label{subsubsec: Testing_results_with_trained_and_finetuned_DeepSC}  
Fig. \ref{fig: Prob_SSM-10000-0906-II} shows the $p(0.1)$ versus $U$ plot exhibited by our relatively wide trained and finetuned DeepSC model tested using 10,000 testing sentences. This plot also demonstrates that our relatively wide trained and finetuned DeepSC model produces semantically irrelevant sentences as the number of MI RFI interferers gets large, also verifying our developed theory \cite{Getu_TWC'24}. Furthermore, comparing Figs. \ref{fig: Prob_SSM-10000-0906-I} and \ref{fig: Prob_SSM-10000-0906-II} confirms that our relatively wide trained and finetuned DeepSC model performs slightly better than our relatively wide trained DeepSC model. As expected, this slight training performance improvement is due to finetuning.  

\subsubsection{Limitations of the Testing Results}
Our recently developed theory predicts that $\lim_{U \to \infty} p(0)=0$. However, our testing results -- depicted in Figs. \ref{fig: Prob_SSM-10000-0906-I} and \ref{fig: Prob_SSM-10000-0906-II} -- show that $p(0.1)$ becomes considerably small when $U$ increases. Although this trend was predicted by our recently developed theory \cite{Getu_TWC'24}, the obtained testing results have limitations due to the following factors: $i)$ The semantic similarity assessment limitation of our adopted sentence Transformer; $ii)$ sentence truncation for sequences of sentences exceeding 30 during training/testing (to alleviate an OOM error we frequently experience); and $iii)$ the training/validation accuracy limit that repeatedly emerges during our training campaign. 
                                              
\section{Concluding Summary and Research Outlook}
\label{sec: summary_and_outlook}
This empirical work studied the impact of interference on a text SemCom system dubbed DeepSC. Specifically, we carried out the training of DeepSC followed by its testing with and without MI RFI using a standard SMT and NMT dataset named Europarl. Using training, validation, and testing sets tokenized and vectorized from Europarl, we trained the DeepSC architecture in Keras 2.9 with TensorFlow 2.9 as a backend, and tested it in the presence and absence of Gaussian MI RFI received over Rayleigh fading channels. For this testing setting, the results obtained using our relatively wide trained DeepSC model and the relatively wide trained and finetuned DeepSC model demonstrated that DeepSC produces semantically irrelevant sentences as the number of Gaussian RFI emitters becomes very large, consistent with our recently developed theory in \cite{Getu_TWC'24}. Accordingly, a fundamental 6G design paradigm for IR$^2$ SemCom is needed, and our (generic) lifelong DL-based IR$^2$ SemCom system \cite[Fig. 2]{Getu_TWC'24} could be the beginning. 

Informed by our multidisciplinary theoretical and empirical research on DL, NLP, NMT, and SemCom, this paper documented extensive details on the steps regarding the training of DeepSC and its testing with and without interference, closing the existing knowledge gap that may have hindered the development of many text SemCom systems. 

\section*{Acknowledgment} 
We gratefully acknowledge the Digital Research Alliance of Canada \cite{DRAC_address} (formerly Compute Canada) for computational support through the Béluga and Graham GPU clusters.

\balance

% Generated by IEEEtran.bst, version: 1.14 (2015/08/26)

%\bibliographystyle{IEEEtran}
%\bibliography{Teref}

%\begin{IEEEbiography}{Tilahun Melkamu Getu}(SM'12)
%Biography text here.
%\end{IEEEbiography}

% if you will not have a photo at all:
%\begin{IEEEbiographynophoto}{Wessam Ajib}
% Biography text here.
% \end{IEEEbiographynophoto}

% \begin{IEEEbiographynophoto}{Omar A. Yeste-Ojeda}
% Biography text here.
% \end{IEEEbiographynophoto}
\end{document}